\documentclass{eas}
\usepackage{graphicx}
\TitreGlobal{The Future Astronuclear Physics}
\begin{document}

\newcommand{\lsim}{\mbox{\raisebox{0.3ex}{$<$}\raisebox{-0.7ex}
{\hspace*{-0.8em}$\sim$}}\/}
\newcommand{\gsim}{\mbox{\raisebox{0.3ex}{$>$}\raisebox{-0.7ex}
{\hspace*{-0.8em}$\sim$}}\/}

\title{Required precision of mass and half-life measurements for
r-process nuclei planned at future RI-beam facilities}
\runningtitle{Motizuki \etal: Required experimental precision
for r-process nuclei}
\author{Y. Motizuki}\address{RIKEN, Hirosawa 2-1, Wako 351-0198 Japan}
\author{T. Tachibana}\address{Senior High School of Waseda University, 
Kami-Shakujii 3-31-1,
Nerima-ku, Tokyo 177-0044 Japan}
\secondaddress{Advanced Research Institute for Science and Engineering, 
Waseda University,
Okubo 3-4-1, Shinjuku-ku, Tokyo 169-8555, Japan}
\author{S. Goriely}\address{Institut d'Astronomie et d'Astrophysique,
Universite Libre de Bruxelles, CP226 Brussels, BEL-1050 Belgium}
\author{H. Koura}
\sameaddress{1,3}
%
%
\begin{abstract}
In order to understand the r-process nucleosynthesis, we suggest
precision required for mass and $\beta$-decay
half-life measurements planned at future RI-beam facilities.
To satisfy a simple requirement that we put on nuclear model predictions,
it is concluded that the detectors for the mass measurements
must have a precision of 1$\sigma$ \lsim 250 keV,
and that the detectors for the half-life measurements
demand a precision of 1$\sigma$ \lsim 0.15 ms.
Both the above precisions are required at the neutron richness of $A/Z$ = 3.0
at the $N$=82 shell closure and $A/Z$ = 2.9 at the $N$=50 shell closure.
For the doubly magic nuclide $^{78}$Ni, 
a precision of 1$\sigma$ \lsim 300 keV and 
1$\sigma$ \lsim 5 ms are required, respectively, 
for mass and half-life measurements.
This analysis aims to provide a first rough guide for ongoing detector developments.
\end{abstract}
\maketitle

\section{Introduction}
 The r-process nucleosynthesis is called for to explain the origin of about half the
elements heavier than iron observed in nature. Its astrophysical origin remains a
mystery. The r-process is one of the most complex nucleosynthesis process to
explore because of the numerous difficulties still affecting the description of both the
explosive astrophysical conditions believed to host the process and the nuclear
properties of the exotic neutron-rich nuclei involved. From the nuclear physics point of
view, the major difficulty lies in the determination of nuclear data for the thousands
of nuclei far from the $\beta$-stability, for which essentially no experimental data
exist nowadays. These concern mainly nuclear structure properties, $\beta$-decays,
neutron captures, photodisintegrations as well as fission processes. In
particular, mass predictions for neutron-rich nuclei play a key role since they affect
all the nuclear quantities of relevance in the r-process, \ie\ the $\beta$-decay, 
neutron capture and photodisintegration rates, as well as the fission probabilities.

Future RI-beam facilities, which are now under construction or planning,
place their first priority to measure masses and half-lives of
neutron-rich nuclei which have not been observed yet and are relevant to
the r-process studies. 
In the coming experiments, it is clearly meaningful to measure 
such masses and half-lives of
unknown neutron-rich nuclei. In addition, we emphasize that
information on their experimental errors is crucial
to promote theoretical studies on mass and $\beta$-decay half-life.
The present paper aims at guiding such future experiment
in defining {\em how far} from the stability line and {\em how much precisely} these
physical quantities should be measured. 

Different approaches can be followed to answer such questions. However, one major
fact that should be kept in mind is that the r-process astrophysical site remains
totally unknown to date. Although the solar system signature clearly shows that the
nuclear mechanisms responsible for the production of r-process 
nuclei concern neutron captures
and beta-decay in the exotic neutron-rich region, no astrophysics model can nowadays
consistently predict the neutron densities required for a successful r-process. 
The ``hot bubble'' scenario or the postexplosion outflows 
expected from protoneutron stars in the seconds 
after successful core-collapse supernovae 
are thought to be a likely candidate site for the r-process
(e.g., Meyer \etal~1992; Woosley \etal~1994).
However, recent models of 
spherical ``neutrino-driven winds'' from protoneutron stars
(e.g. Takahashi \etal~1994; Thompson \etal~2001) 
fail to produce robust r-process nucleosynthesis
up to and beyond the third ($A \approx 195$)
r-process peak for ``canonical'' neutron stars
with $M$ = 1.4 $M_\odot$ and $R$ = 10 km.
The other proposed sites include such scenarios as
``neutron star mergers'' (Freiburghaus \etal~1999),
weak r-process by the shock processing of the helium and/or carbon shells of 
core-collapse supernovae (Truran \& Cowan 2000),  
magnetic protoneutron star winds (Thompson 2003),
prompt explosions from collapsing O-Ne-Mg cores (Wanajo \etal~2004), 
or even interestingly, some settings with rapid ejection of high-entropy but 
nearly symmetric matter to produce the r-process nuclei without excess neutrons
(Meyer 2002). 
Each of them, however, faces severe
problems and cannot at the present time explain the production and
galactic enrichment of the r-process nuclei observed in the Universe.
Moreover, recent observational studies (e.g. Sneden \etal~2000)
of the relative abundance pattern of the r-process elements in 
very metal poor stars 
and also analysis (e.g., Wasserburg, Busso \& Gallino 1996) 
based on the isotopic abundances for the early solar system measured in meteorites
have suggested that different r-process sites are responsible for
the lighter ($A$ \lsim 135-140) and heavier ($A$ \gsim 135-140) r-process nuclei. 
This makes the determination of the physical characteristics 
for the r-process environment further complicated.  

For the above reason,
it remains extremely difficult to estimate the precision required for mass and
$\beta$-decay half-life measurements on the basis of r-process abundance calculations.
In order to answer the objective questions treated in the present paper, 
we have therefore chosen to consider criteria
independently of any ``realistic'' astrophysics calculations. 
Even when the future experiments are performed, 
it is clear that theoretical predictions will still have to fill the
experimental gaps for the thousands of nuclear data required in r-process
simulations. In the first step, these future measurements will therefore mainly  help in
improving the theoretical models by constraining them further on nuclei closer to the
one involved by the r-process, or even directly involved. They might bring new insights
on nuclear physics phenomena at large neutron excesses as well as improve the present
parametrizations of mass formulas. Although most of the recent mass formulas show fits
to experimental masses of similar quality, the mass extrapolations far from the valley of
$\beta$-stability can differ from each other quite significantly (for a recent review,
see Lunney \etal~2003). We have therefore chosen to estimate the nuclei to be involved
and the required precision of future measurements by considering arguments
on simple astrophysics considerations and existing nuclear model predictions as
explained below.

When the future experiments supply with information on new masses and half-lives
with a reasonable precision,
model predictions will tend to converge
if their parameters are updated to fit the newly measured masses. In this regard, mass
formula studies would not benefit if the experimental
errors do not resolve the differences between the model
predictions for the most exotic neutron-rich nuclei accessible (or ideally
directly involved in the r-process nuclear flow).  Accordingly, as a first rough guide
for the required precision in detector developments, we put a rather simple requirement
to mass and half-life measurements as follows: Experimental errors subsidiary to the
r-process nuclei need to be less than {\em half} the difference between the masses
(or half-lives) predicted by the different nuclear models.
We stress that the total length of the error bar obtained in such a procedure
corresponds to $\pm 3\sigma$.  In the following, we discuss such a criterion
on neutron-rich nuclei at the $N$=50 and $N$=82 shell closures.
These regions are expected to become accessible in near-future experiments and are known to
be of first importance in the development of nuclear structure studies, as well as in
our understanding of the r-process nucleosynthesis.

\section{Properties of Considered Mass Formulas}

We consider here three mass formulas, known as HFB-2  (Goriely \etal \, 2002; Samyn
\etal \, 2001), FRDM  (\cite{FRDM}), and KUTY (\cite{KUTY}), available for a wide-range
use in the nuclear chart and hence at this moment appropriate for r-process abundance
calculations. The three mass formulas predict the 2135 measured masses with a
root-mean-square deviation of about $\sim$680 keV (see \cite{LPT03}), 
although they were derived from quite different leading principles.
The HFB-2 model is taken as representative of the microscopic mass formulas recently
derived within the Hartree-Fock-Bogoliubov framework based on an effective nuclear force
of the Skyrme type. On the other hand, the KUTY mass formula corresponds to a
semi-empirical approach making use of an empirical gross term for the  macroscopic
properties of spherical nuclei and spherically-based shell terms for the microscopic
corrections. Here the deviation from the gross properties is explained microscopically
as  shell and deformation effects.
The use of a large number of parameters to describe the single-particle potential
and nuclear gross properties enables
the KUTY model to reproduce relatively well all experimentally known masses as
well as the single-particle energy levels. The FRDM model is also of the semi-empirical
type and was derived from the Finite-Range Droplet Model
for the macroscopic part, and from a deformed single-particle potential for the 
microscopic part.

\begin{figure}[bt]
\begin{center}
\includegraphics[scale=0.35]{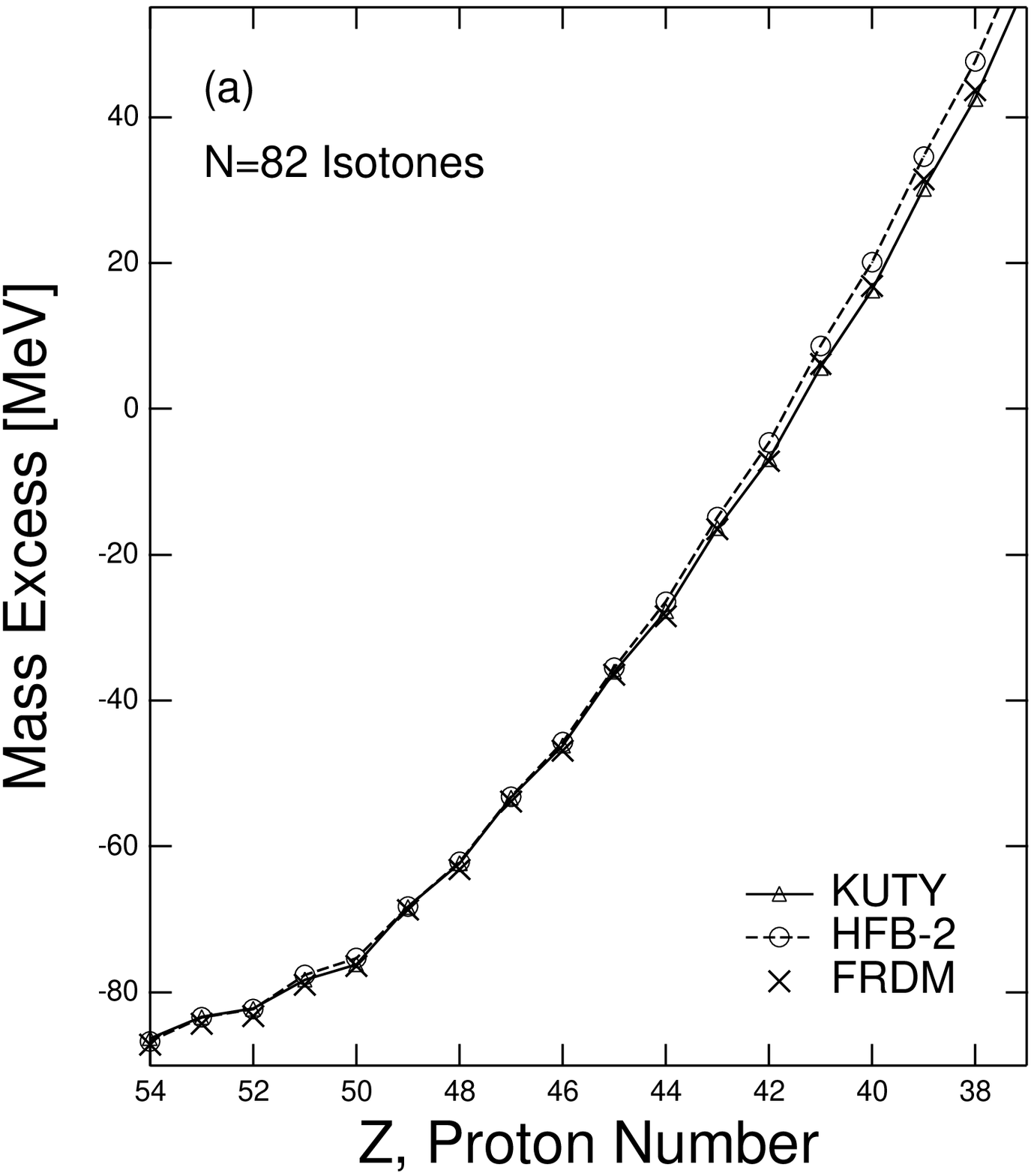}
\qquad
\includegraphics[scale=0.35]{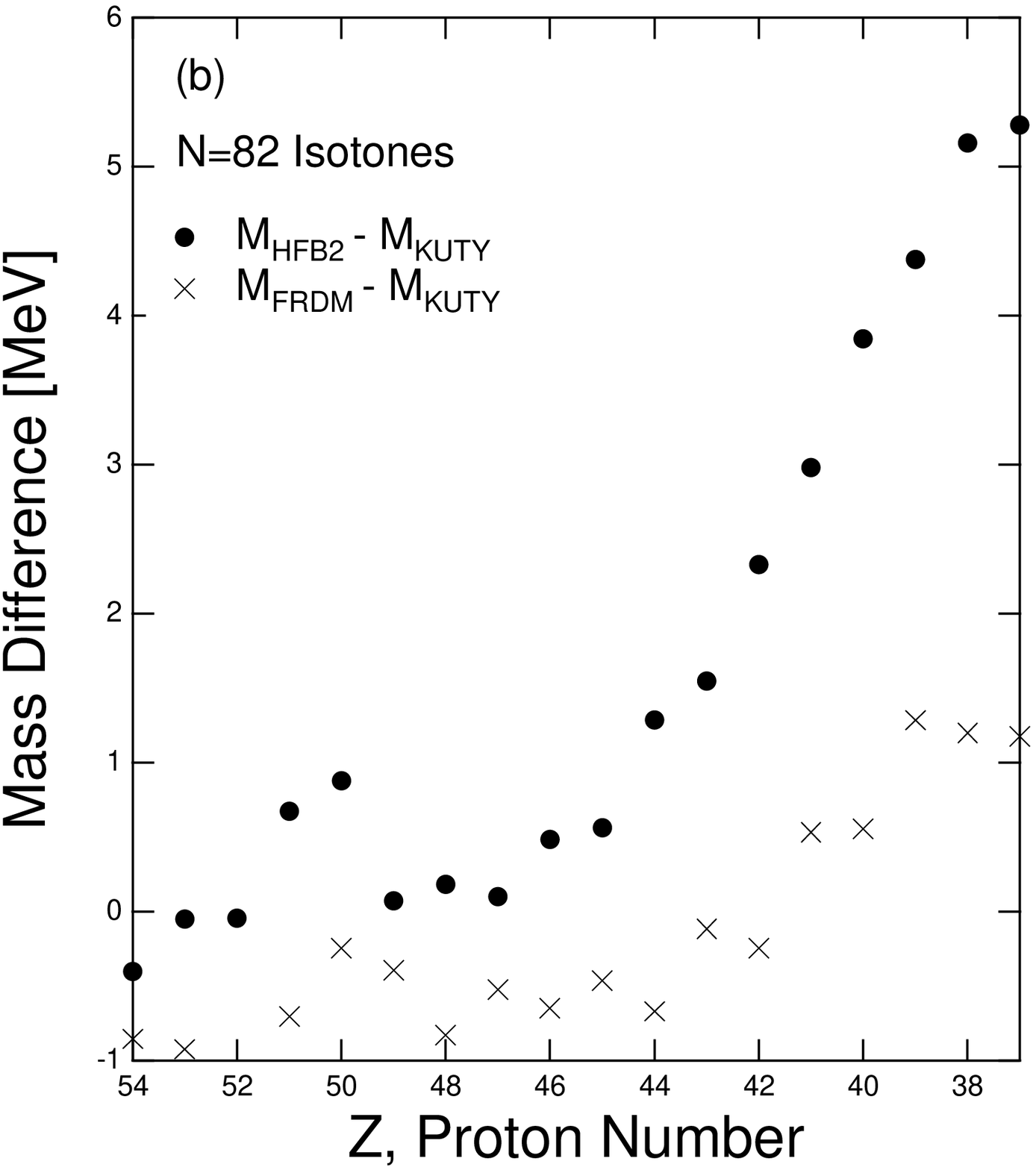}
\end{center}
\caption{(a) A sectional diagram of the $\beta$-stability valley
of its neutron-rich side in the N=82 plane.
The values of mass excess from three theoretical mass formulas are
shown against proton number, Z.
The open circles connected with the dashed line depict the mass excess
from HFB-2, the triangles connected with the solid line from KUTY,
and the crosses from FRDM.
The origin (Z=54, N=82) is the stable $^{136}$Xe while the edge of the abscissa
corresponds to the HFB-2 neutron drip line.
(b) Predicted mass differences are plotted against Z for the N=82
isotones.
The filled circles show the mass difference between HFB-2 and KUTY,
and the crosses between FRDM and KUTY.
The scale of the abscissa is the same as in (a).
}
\label{fig1}
\end{figure}

In Fig.~1(a), we illustrate a global feature of the $\beta$-stability valley
given by the three above-mentioned models.
In the figure a sectional diagram of the neutron-rich side of the valley
is shown for the N=82 isotones.
The microscopic mass formula is seen to give
a steeper slope of the $\beta$-stability valley and hence predicts larger
masses compared with those predicted from semi-empirical formulas.
This can be seen in Fig.~1(a) especially for low Z, 
\ie\ at large neutron excesses.

Figure~1(b) depicts
the mass differences between HFB-2 and KUTY and between
FRDM and KUTY for the N=82 isotones.
In particular, it is seen that the mass difference
 between HFB-2 and KUTY is prominent
and increases with decreasing proton numbers, \ie\ when approaching the neutron-drip line.
Namely, both models predict significantly different masses for the neutron-rich nuclei 
far from the stability line. When applied to r-process calculations, such mass differences
inevitably lead to different r-abundance patterns
(Motizuki \etal~2004; Wanajo \etal~2004). 
In the following, we will focus on the two mass formulas, HFB-2 and KUTY, for which the mass
differences are seen to be the most significant ones.

\section{Required Precision for Mass and Half-Life Measurements}

The r-process
is believed to reach the neutron richness of $A/Z \simeq 3$
in dynamical simulations (e.g. Motizuki \etal~2004)
as well as in the simple parametrized site-independent model (e.g. Goriely \& Arnould
1996). For example, within the canonical model prediction making use of the KUTY masses
and the half-lives calculated by the second version of the gross
theory (see below), the r-process path determines
$^{123}$Nb as the polestar neutron-richest nuclide among the N=82 isotones. This result
is obtained assuming the so-called waiting-point approximation at a temperature of $1.5
\times 10^{9}$ K and a neutron density of $10^{24} {\rm cm^{-3}}$ in order to reproduce
the location and width of the $A \approx$ 130 peak observed in the solar
r-process abundance distribution.  The nuclide
$^{123}$Nb is characterized by a ratio of $A/Z=3.0$.
We accordingly analyze the mass differences between the KUTY and HFB-2 
models for the $N$=82 isotones down to $^{123}$Nb.
As seen in Fig.~2(a), the mass difference at $^{123}$Nb is about 3~MeV.
As mentioned in Sect.~1,
the required total error bar $\pm 3\sigma$ is set to {\em half}
the mass difference, so that the detectors for mass measurements
must have a precision of 1$\sigma$ \lsim 250 keV
at the neutron richness of $A/Z$ = 3.0.

\begin{figure}[hbt]
\begin{center}
\includegraphics[scale=0.31]{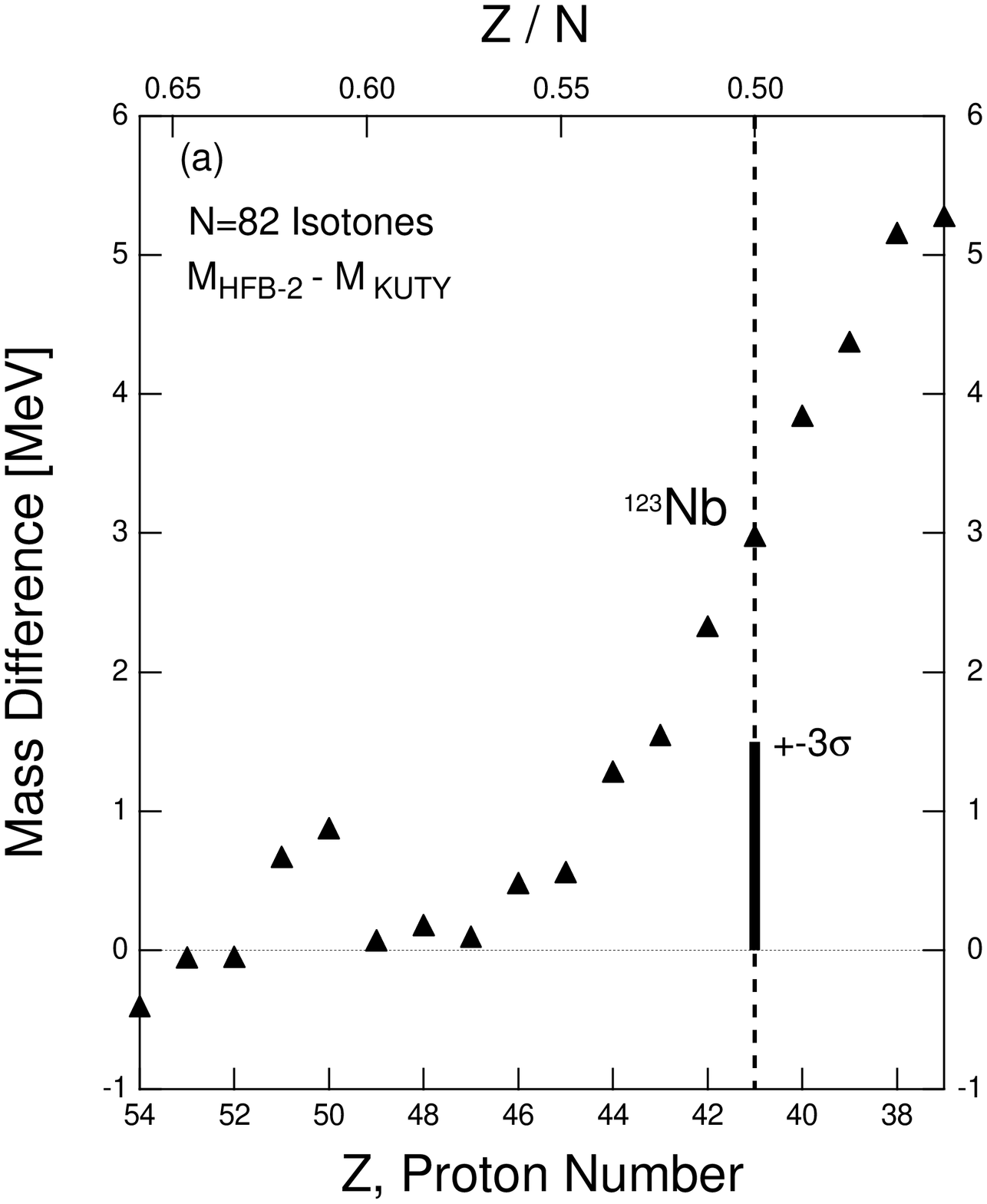}
\qquad
\includegraphics[scale=0.31]{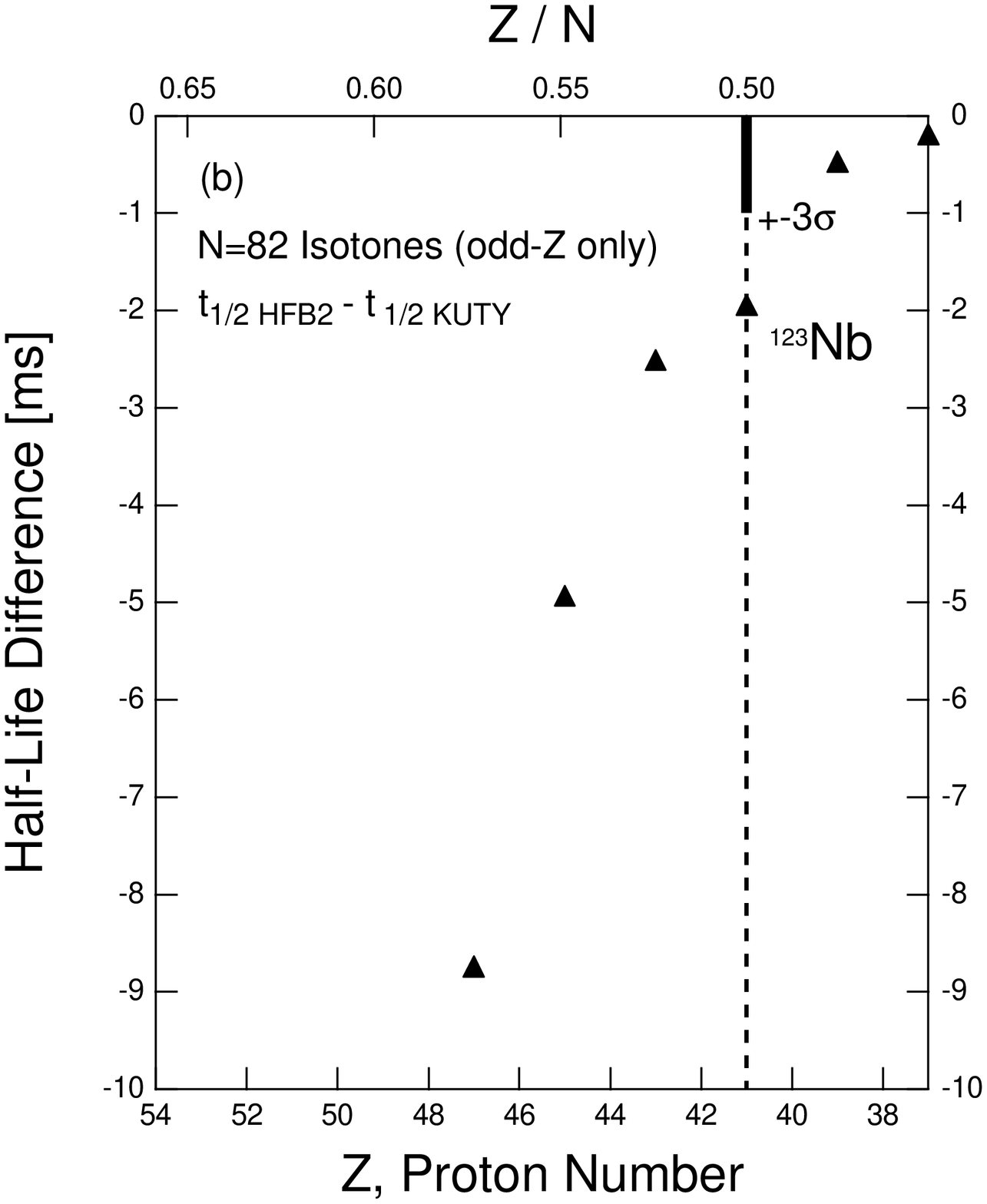}
\end{center}
\caption{(a) Mass difference between the HFB-2 and KUTY predictions
for the $N$=82 isotones is plotted against $Z$.
The vertical dashed line indicates $A/Z = 3.0$,
down to which the r-process is believed to be extended.
(b) Beta-decay half-life difference between the values 
calculated by the second version of the gross theory (GT-2) with HFB-2 masses
and with KUTY for the odd-Z N=82 nuclei.
Both in (a) and (b), the total length of the error bars are indicated
so as to satisfy the requirement explained in Sect.~1.}
\end{figure}

Similarly, we calculate the $\beta$-decay half-life of the N=82 isotones, and
more particularly of $^{123}$Nb, within different approaches to estimate the precision
required in half-life measurements. One of the widely used models used for astrophysics
applications is the second version of the gross theory, known as GT-2 (\cite{T90,T95}). We
apply this model using both the HFB-2 and KUTY $Q_{\beta}$ predictions. Figure~2(b)
shows the half-life difference between both models for the odd-Z N=82 nuclei. We observe
that the half-lives calculated from the microscopic mass formula are shorter than those
from the semi-empirical mass formula. This can be understood by the steeper slope of the
$\beta$-stability valley for the microscopic mass formula: the HFB-2 model
leads to essentially larger $Q_{\beta}$-values than the KUTY model. In particular, for
$^{123}$Nb, the half-life is 3~ms for GT-2 with HFB-2 masses and 5~ms with KUTY. 

However,
the uncertainties in $\beta$-decay predictions stem not only from mass predictions, but
also from the theoretical model used to describe the weak interaction. Mean field and
shell models have been applied in recent years to the calculation of the
$\beta$-decay rates of nuclei of astrophysics interest. In the
particular case of $^{123}$Nb, these models predict a half-life of about 4~ms for the DF3
density functional plus continuum QRPA approximation of Borzov (2003) including only the
allowed transitions and about 3~ms if the first forbidden transitions are also included.
A shorter half-life of about 2~ms is obtained by the shell model of Martinez-Pinedo \&
Langanke (1999).  Considering such half-life differences for
$^{123}$Nb, we find that 1$\sigma$ \lsim 0.15 ms at $A/Z=3.0$ is required for the
half-life measurements.

A similar procedure can be followed in the $N$=50 region.
We find that similar precisions
(1$\sigma$ \lsim 250 keV for masses and
1$\sigma$ \lsim 0.15 ms for half-lives) are required from GT-2 calculations 
at $A/Z$ = 2.9 on the $N$=50 shell closure, \ie, for $^{76}$Fe.
However, if we consider the doubly
magic nuclide $^{78}$Ni which has been observed but for which the mass and the half-life
remains experimentally unknown, the same criterion leads us to a precision of 1$\sigma$
\lsim 300 keV for mass and of 1$\sigma$ \lsim 5~ms for half-life measurements.

\section{Summary and Feasibility}
We have derived the required precision of
1$\sigma$ \lsim 250 keV and 1$\sigma$ \lsim 0.15 ms, respectively,
for mass and half-life measurements
at the neutron richness of $A/Z$ = 3.0 at the $N$=82 shell closure and
at the $A/Z$ = 2.9 at the $N$=50 shell closure. 
For the doubly magic nuclide $^{78}$Ni, we have found that the detectors
must have a precision of 1$\sigma$
\lsim 300 keV for mass and of 1$\sigma$ \lsim 5~ms for half-life measurements.
Note that not only statistical but also
systematic errors should be included in the above discussion. 
It should also be kept in mind that the precision estimate presented here is based on
simple arguments due to our ignorance of the astrophysical site for the r-process.
Future
development in nucleosynthesis models (Takahashi, this volume) will hopefully bring new
insight on the nuclear flow followed by the r-process and consequently on the nuclei
involved and the major nuclear quantities of relevance.

Experiments at RIKEN RI-Beam Factory will start in 2007.
Here RI-beams are planned to be produced by fragmentation and uranium 
fission methods.
The intensity of the RI-beams will be strong enough to reach $^{78}$Ni
($A/Z$=2.8) 
and $^{76}$Fe ($A/Z$=2.9) at the $N=50$ shell closure 
to measure these masses and half-lives with the suggested precisions.
However, the expectations of the RI-beam intensity
created with the {\em fragmentation} method at present
come down to one particle
per $10^{5}$ sec at the $A/Z$=3.0 ($^{123}$Nb) region 
at the $N=82$ shell closure.
This means that the measurements with the required precisions might be 
difficult for the present technology:
It is indispensable to contrive new type of detectors to overcome this 
difficulty.

Future measurements with better precision are strongly encouraged in order to develop
theories of nuclear masses and half-lives.
Progress in these theories and above all in microscopic approaches, as well as further
developments of astrophysics models will help us to solve the long-standing mystery that
the r-process nucleosynthesis still represents.

\hspace{2em}

We would like to thank K. Takahashi for useful comments.
Y.M. would like to acknowledge Y. Ishida, T. Suda,
and Y. Yano for information on experimental status at the RI-Beam Factory. S.G. is FNRS
Research Associate.


\end{document}